\documentstyle[preprint,aps]{revtex}
\input{psfig}
\newcommand{\omnull}[2]{\mbox{$ \Omega^0_{\vec{#1},\, \vec{#2}}$}}
\newcommand{\omeins}{\mbox{$ \Omega^1 $}}
\newcommand{\omzwei}[2]{\mbox{$ \Omega^2_{\vec{#1},\, \vec{#2}}$}}
\newcommand{\bra}[1]{\mbox{$\langle #1|$}}
\newcommand{\ket}[1]{\mbox{$|#1\rangle$}}
\newcommand{\Id}{{\rm 1\hspace{-0.75ex}1}}

\begin{document}
\draft
\preprint{MKPH-T-96-21}
\title{{ \bf Coherent eta photoproduction on the deuteron in the $S_{11}$ 
resonance region}}

\author{E.\ Breitmoser
 and H.\ Arenh\"ovel}
\address{
Institut f\"{u}r Kernphysik, Johannes Gutenberg-Universit\"{a}t,
       D-55099 Mainz, Germany}
\maketitle

\begin{abstract}
Coherent eta photoproduction on the deuteron is studied in the $S_{11}$ 
resonance region neglecting eta rescattering and two-body processes. For the 
elementary reaction on the nucleon, we have considered the dominant 
$S_{11}(1535)$ resonance and as background the nucleon pole terms as well 
as $\rho$ and $\omega$ meson exchange. We have studied the influence of 
different 
choices for the neutron resonance amplitude, different prescriptions for 
fixing the invariant mass of the resonance amplitude and different methods 
for deriving the elementary production amplitude for an arbitrary reference 
frame.
\end{abstract}

\section{Introduction}
In recent years, photo- and electroproduction 
of $\eta$ mesons on nucleons and nuclei have been studied intensively 
\cite{TiB95,BeT95,SaF95a}. This 
renewed interest has been triggered primarily by the significant 
improvement on the quality of experimental data obtained with the new 
generation of high-duty cycle electron accelerators like, e.g., ELSA in Bonn 
and MAMI in Mainz \cite{Wil93,KrA95a,PrA95}. The special interest in the 
electromagnetic production of $\eta$ mesons on the nucleon is based on the 
fact that, being an isoscalar meson, it constitutes a selective filter for 
isospin $I=\frac{1}{2}$ nucleon resonances $N^{*}$.  
Furthermore, the e.m.\ $\eta$ production is dominated by the intermediate 
excitation of the $S_{11}$(1535) resonance. Thus this reaction is an ideal 
tool for investigating the characteristic features of this resonance, which 
usually is difficult to separate  from the other resonances because of their 
large widths. For example, one can study its electromagnetic transition 
multipoles and its decay channels, thus providing a good test for hadron 
models. 

Eta photoproduction on the deuteron is of particular interest since first it 
allows to study this reaction on a bound nucleon in the simplest nuclear 
environment, and second, it provides important information on this reaction 
on the neutron. With respect to the latter case, the deuteron is considered 
as a neutron target assuming that binding effects can be 
neglected. The first data for coherent $\eta$ photoproduction on the 
deuteron were obtained almost thirty years 
ago by Anderson and Prepost \cite{AnP69}. Their results for the weighting of 
isoscalar and isovector amplitude of the $S_{11}$(1535) resonance 
were at variance 
with quark model predictions and data analysis for pion photo production
on the nucleon in the region of the $S_{11}$ resonance. 
Only with a large isovector 
amplitude, the theory could explain the data assuming the validity of the 
impulse approximation \cite{KrL76}. Other explanations were proposed by 
Hoshi et al. \cite{HoH79} by invoking a dominance of $\eta$ rescattering on 
the other nucleon or a large influence of the $P_{11}(1440)$ resonance. 
The latter seems to be ruled out at present. Furthermore, their treatment 
of rescattering is very approximate and a more recent evaluation of 
rescattering effects, although still approximate, gave a much smaller effect 
\cite{HaR89}. Therefore, improved calculations are urgently needed. Very 
recently, rescattering has been considered in \cite{KaT95} in the 
near threshold region showing only a small effect. 
It is clear that new data will offer the
chance to investigate this problem in greater detail. With respect to new 
experimental data, only upper limits exist at present for the reaction 
on the deuteron \cite{Beu94,KrA95b} which, 
however, are already at variance with the old data.  

The aim of the present work is to initiate a systematic study of the 
$d(\gamma,\pi^0)d$ 
reaction on the deuteron from threshold through the $S_{11}$ resonance. 
As a first step, we will restrict 
ourselves to the plane wave impulse approximation (PWIA) in order to study 
details of the elementary reaction amplitude with respect to the yet 
unknown neutron properties and to study different ways of implementing the 
elementary amplitude in a bound system. 

Among the phenomenological models describing existing data for the elementary 
process on the proton for
energies below 1 GeV in a satisfactory manner, there are the isobar model 
\cite{Hic73}, the coupled channel model \cite{BeT91} and the effective 
Lagrangian approach \cite{BeM94,Kno94,SaF95}. 
We have chosen the Langrangian approach of \cite{BeM94,Kno94} because the
resonance and background terms are treated on the same level and the number
of unknown parameters is much less than that for the two other models. 
Near threshold the
resonances $P_{11}$(1440), $D_{13}$(1520) and $S_{11}$(1535) are
likely to contribute. Although the $P_{11}$(1440) is below the threshold, 
it can influence the reaction because of its large
width. Little is known on the decay widths of the resonances $P_{11}$(1440) and
$D_{13}$(1520) to $\eta N$ and they are not listed in \cite{PDG94}. 
In \cite{PDG92}, the branching ratio of the $D_{13}$(1520) into this channel 
is listed to be  $0.1\%$. Furthermore, the experiment of Krusche et al.\ 
\cite{KrA95a} did not give any evidence for the 
influence of the $P_{11}$(1440), whereas the $D_{13}$(1520) was seen for the
first time. Nevertheless, it was shown that the influence of the
$D_{13}$(1520) to the total cross section was negligible. For these
reasons, it is legitimate to consider in the present study 
only the dominant $S_{11}$ resonance. 
In addition we include background terms such as nucleon pole terms 
and the exchange of the vector mesons $\rho$ and $\omega$. 
Apart from the resonance, also
the strength and nature (pseudoscalar or pseudovector) of the 
$\eta NN$-coupling is widely unknown. 

When implementing the elementary reaction amplitude into the deuteron, one 
has to face several problems which are analogous to coherent 
$\pi$ photoproduction on the deuteron. First of all, the energy of 
the struck nucleon on which the reaction takes place is not well defined in 
a bound nonrelativistic system. Consequently, the invariant mass of the 
$\gamma N$ subsystem of the elementary reaction, on which the elementary 
$t$-matrix depends, is not well defined. In particular, the invariant 
mass determines the decay width of the resonance to which
the resulting cross section is very sensitive. Secondly, 
the elementary reaction amplitude, which usually is given 
in the $\gamma N$ c.m.\ frame, has to be transformed to an arbitrary 
reference frame. This may be done either by a Lorentz boost of all four 
momenta on which the elementary process depends or by calculating the 
$t$-matrix with respect to an arbitrary frame right from 
the beginning. The last method is more general because one does 
not loose any terms which vanish in the $\gamma N$ c.m.\ frame. 
But in both cases one faces 
again the problem of assigning an energy to the bound nucleon. 
We will compare both methods for the contributions of vector meson 
exchange in order to check to what extent the differences matter. 
   
The paper is organized as follows: In the next section we present the 
model for the elementary reaction, and fix all parameters for the 
process on the proton. 
Then we discuss in Sect.\ III the theoretical changes that have to be 
considered for the process on the deuteron. In Sect.\ IV we present 
our results and compare them with existing experimental data and 
other theoretical calculations. Finally, we give a brief summary and 
an outlook. Some calculational details are collected in an appendix. 

\section{The elementary process $\gamma + N \rightarrow \eta + N$}
\label{elprocess}
In this section, we will briefly review the elementary reaction in order to 
introduce the model and to fix as many model parameters as possible for 
the implementation of the process into the deuteron. 
The Mandelstam variables of the reaction $\gamma (k) + N(p) \rightarrow 
\eta (q) + N(p')$ are defined as usual 
\begin{eqnarray}
s & = & (k_\mu+p_\mu)^{2}=(q_\mu+p_\mu^{\,\prime})^{2}, \\
t & = & (k_\mu-q_\mu)^{2}=(p_\mu-p_\mu^{\,\prime})^{2}, \\
u & = & (k_\mu-p_\mu^{\,\prime})^{2}=(q_\mu-p_\mu)^{2}, 
\end{eqnarray}
where $p_\mu$ ($p_\mu^{\,\prime}$) denotes the four momentum of the incoming 
(outgoing) nucleon, $k_\mu$ and $q_\mu$ the momenta of photon and eta meson, 
respectively. The energies of the participating particles are given by 
\begin{eqnarray}
k_{0} & = & |\vec{k}|, \\
\omega_{q} & = & \sqrt{m_{\eta}^{2}+\vec{q}\,^{2}}\,, \\ 
E_{p^{(\prime)}} & = & \sqrt{M^{2}+\vec{p}\,^{(\prime)\,2}}\,.
\end{eqnarray}
The invariant mass of the $\gamma N$ system is given as function of the 
photon lab energy $E^{Lab}_{\gamma}$ by 
\begin{equation}\label{equ:inv}
W_{\gamma N}=\sqrt{s}=\sqrt{M^{2}+2E^{Lab}_{\gamma} M}\,,
\end{equation}
and the absolute values of the three momenta in the c.m.\ system by 
\begin{eqnarray}
k & = & \frac{s-M^{2}}{2\sqrt{s}}, \\ \label{equ:q}
q & = & \frac{1}{2\sqrt{s}}
\sqrt{[s-(M+m_{\eta})^{2}][s-(M-m_{\eta})^{2}]}\,.
\end{eqnarray}
Nucleon and eta masses are denoted by $M$ and $m_\eta$, respectively. 
In the c.m.\ system, the unpolarized differential cross section is given by 
\begin{equation}
\frac{d \sigma}{d \Omega_{\eta}^{c.m.}}=\frac{1}{6}\frac{1}{16\pi^{2}} 
\frac{q}{k} 
\frac{E_k E_q}
{W_{\gamma N}^{2}}
\sum_{m^{\prime} m \mu}
|\tau_{m^{\prime} m \mu}(W_{\gamma N},\theta)|^{2}\,,
\end{equation}
where $\tau_{m^{\prime} m \mu}$ denotes the $t$-matrix element for initial 
and final nucleon spin projections $m$ and $m'$, respectively, and photon 
polarization $\mu$ 
\begin{equation}
\tau_{m^{\prime} m \mu}(W_{\gamma N},\, \theta) = \langle m'|
t_{\gamma \eta}(W_{\gamma N})|m \mu
\rangle\,,
\end{equation}
and $\theta$ the angle of the outgoing $\eta$ meson with respect to the 
incoming photon momentum. 
Here we have assumed covariant state normalization, i.e., 
\begin{equation}
\langle p^{\,\prime}\,|p\rangle=(2\pi)^3\frac{E_p}{M}\delta (\vec 
p^{\,\prime}-\vec p)
\end{equation}
for fermions and
\begin{equation}
\langle p^{\,\prime}\,|p\rangle=(2\pi)^3 2\omega_p\delta (\vec
p^{\,\prime}-\vec p)
\end{equation}
for bosons. 

\begin{figure}
\centerline{\psfig{figure=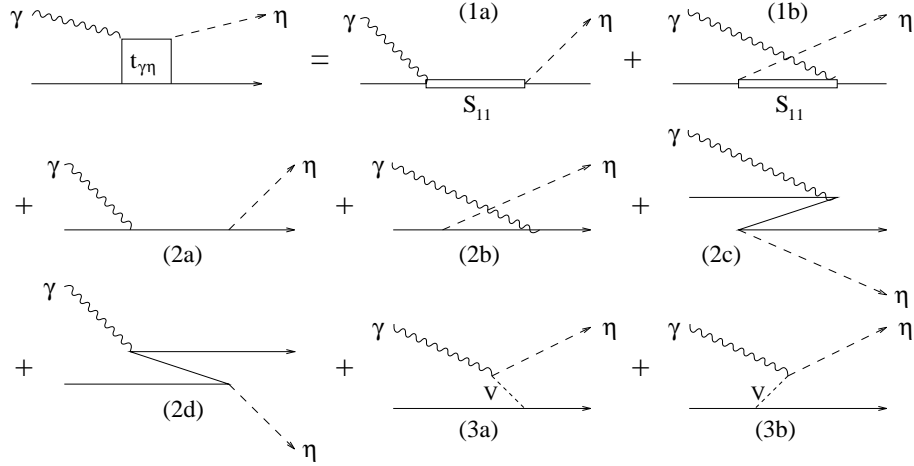,width=12cm,angle=0}}
\vspace*{.5cm}
\caption{Time ordered diagrams of all contributions: 
$S_{11}$ resonance in $s$- (1a) and $u$-channel (1b),  
nucleon pole terms in $s$- (2a) and $u$-channel (2b) with z-graphs (2c) and 
(2d), and $t$-channel vector meson (V) exchange (3a) and (3b). 
\label{fig:a}}
\end{figure}

Besides the dominant $S_{11}$ resonance in the $s$-channel, we consider 
for the $t$-matrix the $S_{11}$ in the $u$-channel, the nucleon pole terms 
in the $s$- and $u$-channels, as well as $\rho$ and $\omega$ exchange in the
$t$-channel as background contributions. The corresponding diagrams are 
shown in Fig.\ \ref{fig:a}. 
All vertices are calculated from the following effective Lagrangians 
\cite{BeM94}
\begin{eqnarray}
{\cal L}_{\eta NN}^{PS} & = & -i g_{\eta} \bar\Psi \gamma_{5} \Psi \phi_{\eta}
\,,\\
{\cal L}_{\gamma NN} & = & -e \bar\Psi \gamma_{\mu} \frac{1+\tau_{0}}{2} 
\Psi A^{\mu}-\frac{e}{4 M} \bar\Psi (\kappa^{s}+\kappa^{\nu} \tau_{0}) 
\sigma_{\mu \nu} \Psi F^{\mu \nu} 
, \\
{\cal L}_{VNN} & = & - g_{v} \bar\Psi \gamma_{\mu} \Psi V^{\mu}- \frac{g_{t}}
{4 M} \bar\Psi \sigma_{\mu \nu} \Psi V^{\mu\nu}\,, \label{lvnn}\\ 
\label{equ:he}
{\cal L}_{V\eta \gamma} & = & \frac{e \lambda}{4 m_{\eta}} \varepsilon_{\mu\nu
\lambda\sigma} F^{\mu \nu} V^{\lambda 
\sigma} \phi_{\eta}\,, \\  
{\cal L}_{\eta NS_{11}} & = & -i g_{\eta N S_{11}} \bar\Psi \Psi_{S_{11}} \phi_{\eta}
+h.c\,, \\
{\cal L}_{\gamma NS_{11}} & = & -\frac{e}{2 (M_{S_{11}}+M)} \bar\Psi_{S_{11}} (\kappa^{s}_{
res}
+\kappa^{v}_{S_{11}} \tau_{0}) \gamma_{5} \sigma_{\mu \nu} \Psi F^{\mu \nu}
+h.c. \, .\label{lgns}
\end{eqnarray} 
Here, $\Psi$ and $\Psi_{S_{11}}$ describe the fermion fields of the nucleon 
and the $S_{11}$ resonance, respectively, $V^{\mu}$
the vector meson field, $\phi_{\eta}$ the pseudoscalar eta meson 
field,  and $A^{\mu}$ the photon field with its field tensor $F^{\mu \nu}$. 
The vector meson field tensor is analogously 
defined by $V^{\mu\nu}=\partial^{\mu}V^{\nu}-\partial^{\nu}V^{\mu}$.
The $S_{11}$ mass is denoted by $M_{S_{11}}$ and the isoscalar (isovector) 
anomalous magnetic moment of the nucleon by $\kappa^{s}$ $(\kappa^{v})$. 

For the the $\eta NN$-vertex we have chosen the pseudoscalar coupling 
\cite{KrA95a,TiB94}. Values of the coupling constant range 
usually between 0 and 7 \cite{BeM94,TiB94}. Fitting the model to 
the data, we find as best choice $\frac{g_{\eta}^{2}}{4 \pi}=0.4$ in 
good agreement with results from \cite{KrA95a} and with a recent 
calculation by Kirchbach and Tiator \cite{KiT96} evaluating an effective 
$\eta NN$-vertex associated with the $a_0(980) \pi N$ triangle diagram. 
The coupling constant $\lambda$ of the $V\eta\gamma$-vertex 
can be determined from the known electromagnetic
decay  $V \rightarrow \gamma \eta $ , with $V=\rho$ or $\omega$. Here 
we use $\Gamma_{\rho\rightarrow \gamma \eta}=57.46$ MeV and $\Gamma_{\omega
\rightarrow \gamma \eta}=3.96$ MeV \cite{PDG92}.
At the hadronic $VNN$-vertex, a phenomenological form factor $F(\vec q_V)$
of dipole type is introduced 
\begin{equation}
F(\vec q_V\,)= \frac{(\Lambda_{V}^2-m_{V}^2)^2}{(\Lambda_{V}^2
+\vec q_V\,^2)^2},
\end{equation}
with  $\vec q_V$ being the momentum of the vector meson and
$\Lambda_{V}$ the cut-off mass. A more detailed description, also for
the coupling constants of the hadronic vertex $g_{v}$ and $g_{t}$, may be found
in \cite{Kno94} and references therein. Table \ref{tab:d} summarizes all 
parameters used for the vector mesons.

The photoexcitation of the $S_{11}$ resonance can be described by the 
helicity amplitudes $A^{p,n}_{1/2}$ (with upper index $p$ referring to the 
proton, $n$ to the neutron) which may be split into isoscalar and isovector 
amplitudes by 
\begin{equation}
A^{p,n}_{1/2}= A^{s}_{1/2} \pm A^{v}_{1/2}\,.
\end{equation}
They determine the resonance parameters $\kappa^{s/v}_{S_{11}}$ in 
(\ref{lgns}) and are related to the helicity amplitudes $A^{s/v}_{1/2}$ by 
\begin{equation}
e\kappa_{S_{11}}^{s/v} 
= \sqrt{\frac{2M(M_{S_{11}}+M)}{M_{S_{11}}-M}} A_{1/2}^{s/v}\,.
\end{equation}
Furthermore, the $\eta N S_{11}$-coupling constant is given by
\begin{equation}
\frac{g_{\eta NS_{11}}^{2}}{4\pi}  =  \frac{2M_{S_{11}}^{2}}{q_{\eta}^{*} 
((M_{S_{11}}+M)^{2}-m_{\eta}^{2})} \Gamma^{0}_{S_{11}} ,
\end{equation}
where $q_{\eta}^{*}$ is the $\eta$ momentum at resonance in the $\eta N$ 
c.m.\ frame, given by (\ref{equ:q}) for $s=M_{S_{11}}^2$,  
and $\Gamma^{0}_{S_{11}}$ denotes the $S_{11}$ decay width.
The resonance mass $M_{S_{11}}$, the helicity 
amplitudes $A^{p,n}_{1/2}$, the hadronic
coupling  $g_{\eta NS_{11}}$ and the decay width $\Gamma^{0}_{S_{11}}$ are only
known within large uncertainties \cite{KrA95a,PDG94}. 
For the resonance, we use the free propagator in the form 
\begin{equation}
\Big[W_{\gamma N}-M_{S_{11}}+\frac{i}{2}\Gamma_{S_{11}}(W_{\gamma N})
\Big]^{-1}\,.
\label{propS11}
\end{equation}
The $S_{11}$ resonance mainly decays into $N\eta$ (50 \%), $N\pi$ (40 \%), 
and $N\pi\pi$ (10 \%) \cite{PDG94}. 
Consequently, one has for the total decay width at resonance position
\begin{eqnarray}
\Gamma_{S_{11}}(M_{S_{11}}) & = & \Gamma_{S_{11} \rightarrow\eta N}+\Gamma_{S_{11}
\rightarrow\pi N}+\Gamma_{S_{11} \rightarrow\pi\pi N}\\ \label{equ:235}
& = & 0.5\Gamma^{0}_{S_{11}}+0.4\Gamma^{0}_{S_{11}}+0.1\Gamma^{0}_{S_{11}}\,.
\end{eqnarray}
In (\ref{propS11}) we already have introduced an energy dependend decay width 
$\Gamma_{S_{11}}(W_{\gamma N})$ depending on the invariant mass $W_{\gamma 
N}$ of the system since it gives a much better description of experimental data 
than using a constant width. For the two leading 
decay channels, we have calculated the corresponding partial decay widths from 
the Lagrangians in first order perturbation theory, whereas 
the three body decay $S_{11} \rightarrow\pi+\pi+N$ is treated on a purely 
phenomenological level keeping it constant above its threshold. Then we have 
for the total width as function of the invariant mass 
\begin{eqnarray} \label{equ:a}
\Gamma_{S_{11}}(W_{\gamma N}) & = & \frac{g_{\eta NS_{11}}^{2} 
M}{2 \pi W_{\gamma N}}
{q}_{\eta}(W_{\gamma N}) \Theta(W_{\gamma N}-W^{th}_{\eta N}) \nonumber \\
& & +\frac{g_
{\pi NS_{11}}^{2} M}{2 \pi W_{\gamma N}} {q}_{\pi}(W_{\gamma N}) 
\Theta(W_{\gamma N}-W^{th}_{\pi N }) \nonumber \\
& & +0.1\Gamma_{o}\Theta(W_{\gamma N}-W^{th}_{\pi\pi N}).
\end{eqnarray}
Here, ${q}_{\eta}(W_{\gamma N})$ is given by (\ref{equ:q}) and 
${q}_{\pi}(W_{\gamma N})$ by the 
analogous equation replacing $m_\eta$ by $m_\pi$. The threshold masses for 
the decay channels are correspondingly denoted by $W^{th}_{\eta N}$, 
$W^{th}_{\pi N}$ and $W^{th}_{\pi\pi N}$. 
The coupling constants $g_{\eta N S_{11}}$ and $g_{\pi  N S_{11}}$ are 
fixed by evaluating the partial decay widths at resonance position. 

\begin{figure}
\centerline{\psfig{figure=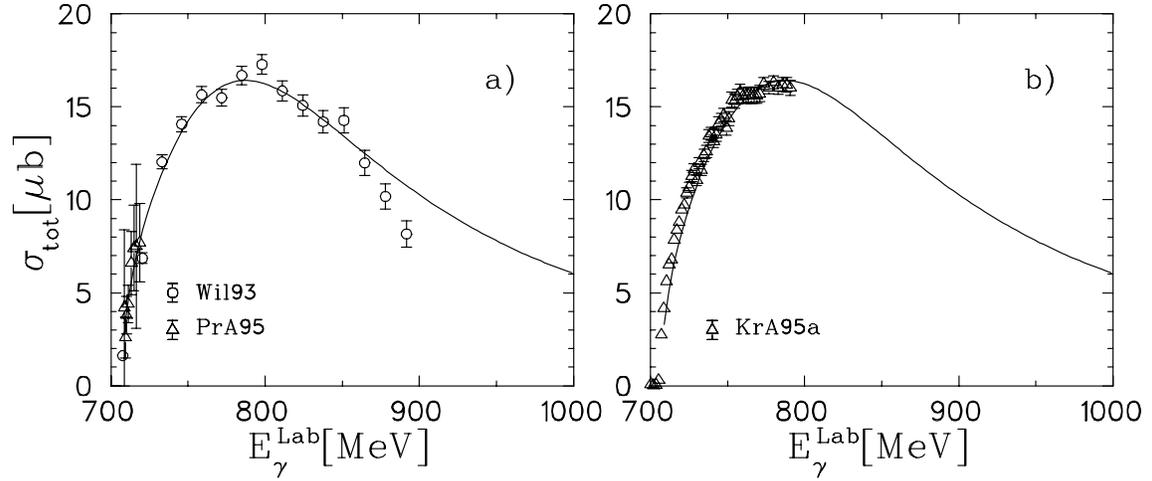,width=15cm,angle=0}}
\vspace*{1cm}
\caption{Total cross section for $\gamma + p \rightarrow \eta +p$ using the 
parameters of Table \protect{\ref{tab:a}} compared to the data of 
\protect{\cite{Wil93}} and \protect{\cite{PrA95}} (left) and 
\protect{\cite{KrA95a}} (right).
\label{tot:etaN}}
\end{figure}

\begin{figure}
\centerline{\psfig{figure=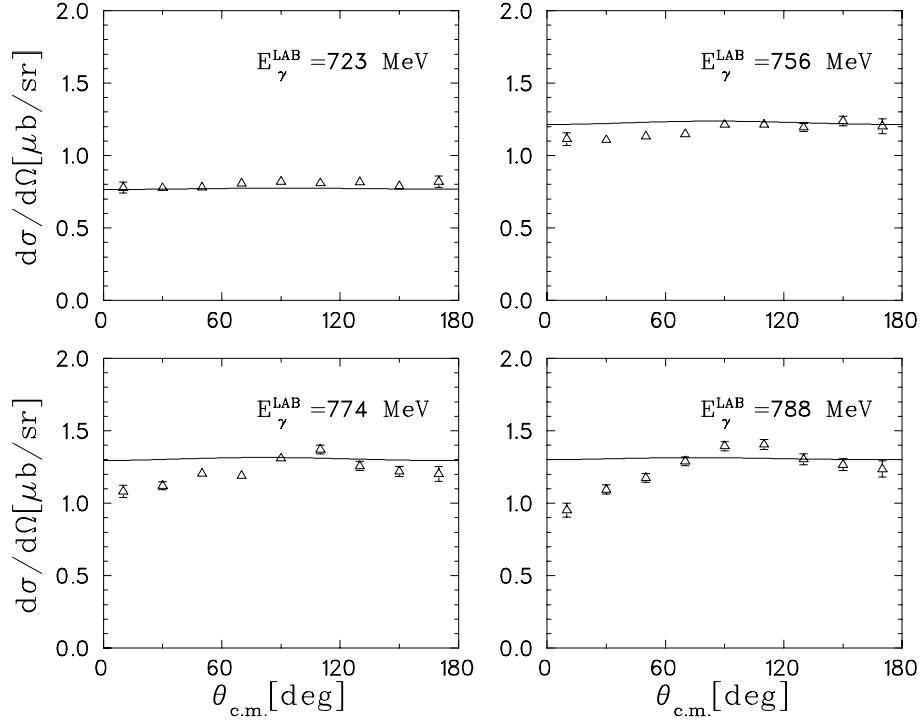,width=12cm,angle=0}}
\vspace*{1cm}
\caption{Differential cross sections for $\gamma + p \rightarrow \eta +p$ 
for four photon energies using the 
parameters of Table \protect{\ref{tab:a}}. Experimental data from 
\protect{\cite{KrA95a}}.
\label{diff:etaN}}
\end{figure}

In Table \ref{tab:a} we list the set of resonance parameters for 
which we obtain the best fit \cite{Bre95} 
to the total cross section data as can be seen in 
Fig.\ \ref{tot:etaN}. Only above 850 MeV, the theoretical cross section 
overestimates the data. The 
differential cross sections calculated with these parameters are shown 
in Fig.\ \ref{diff:etaN}. While at lower energies the angular dependence 
is quite satisfactory, one finds larger deviations at higher energies. 
One reason for this might be the omission of the $D_{13}$ resonance 
\cite{Kno94,TiB94}. However, for the present study this is of no relevance.

\section{The Coherent Process on the Deuteron}\label{deuteron}
Now we turn to the coherent $\eta$ production on the deuteron 
\begin{equation}
\gamma(k_\mu)+d(d_\mu) \rightarrow \eta(q_\mu)+d(d_\mu^{\,\prime})\,.
\end{equation}
The variables in parentheses refer to the corresponding four momenta of the 
particles. We will consider this reaction in the photon-deuteron 
($\gamma d$) c.m.\ 
frame. There we choose the $z$-axis along the photon momentum $(\vec e_z= 
\hat k =\vec k /k)$, the $y$-axis parallel to $\vec k \times \vec q$ and 
the $x$-axis such as to form a right handed system. Thus the outgoing $\eta$ 
meson is described by the spherical angles $\phi=0$ and $\theta$  
with $\cos\theta=\hat{q}\cdot\hat{k}$. 

For the unpolarized differential cross section we obtain in the $\gamma d$
c.m.\ frame \cite{Bre95}
\begin{equation}
\frac{d \sigma}{d \Omega_{\eta}^{c.m.}}=\frac{1}{6}\frac{1}{16\pi^{2}} 
\frac{q}{k} 
\frac{E^{d}_k E^{d}_q}{W_{\gamma d}^{2}}
\sum_{m_{d}m_{d}^{\prime}\mu}
|\tau_{m_{d}m_{d}^{\prime}\mu}(W_{\gamma d},\theta)|^{2} \,,
\end{equation}
with $E^d_p= \sqrt{M_{d}^{2}+\vec p^{\,2}}$. Here we have introduced as 
$t$-matrix
\begin{equation}
\tau_{m_{d}m_{d}^{\prime}\mu}(W_{\gamma d},\theta) = \langle m'_d|
t_{\gamma \eta}^d(W_{\gamma d})|m_d \mu\rangle\,,
\end{equation}
denoting the initial and final deuteron spin projections by $m_d$ and 
$m'_{d}$, respectively, and the photon polarization by $\mu$. Furthermore, 
$W_{\gamma d}$ denotes the invariant mass of the $\gamma d$ system and $k$ 
and $q$ again the photon and $\eta$ momentum, respectively, in the $\gamma 
d$ c.m.\ system. These quantities are given as function of the lab photon 
energy by expressions analogous to (\ref{equ:inv}) through (\ref{equ:q}) 
replacing the nucleon mass $M$ by the deuteron mass $M_d$. 
Here the $\eta$ production threshold is at $E_{\gamma}^{Lab} =629$ MeV, 
which is equivalent to an invariant mass $W_{\gamma d}^{thr}=2424$ MeV. 

\begin{figure}
\centerline{\psfig{figure=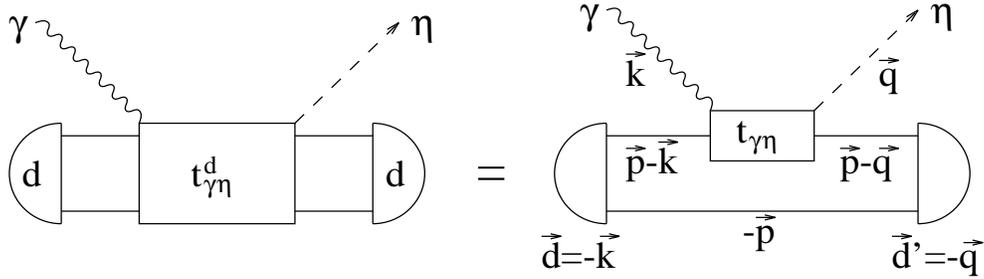,width=13cm,angle=0}}
\vspace*{1cm}
\caption{Diagrammatic representation of $\gamma + d \rightarrow \eta + d$ 
in the impulse approximation with definition of momenta in the $\gamma d$ 
c.m.\ system. 
\label{fig:kk}}
\end{figure}

As noted in the introduction, we restrict our calculation to the impulse
approximation which means that the reaction will take place only 
on one of the two nucleons leaving the other 
as a pure spectator (see the diagram in Fig.\ \ref{fig:kk}).
Consequently, the production operator $t_{\gamma\eta}^d $ for the reaction on 
the deuteron is obtained from the elementary operator $t_{\gamma\eta}$ by
\begin{equation}
t_{\gamma\eta}^{d}=t_{\gamma\eta}^{(1)} \times \Id^{(2)} +\Id^{(1)} \times 
t_{\gamma\eta}^{(2)}\,,
\end{equation}
where the upper index in brackets refers to the nucleon on which the operator 
acts. Off-shell effects will be neglected.
Then the $t$-matrix has the following form
\begin{eqnarray}
\tau_{m_{d}m_{d}^{\prime}\mu} & = &2\sum_{m_s',m_s}
\int d^{3}p \,\psi^{\ast}_{m_s' m^{\prime}_{d}}
\Big(\vec{p}-\frac{\vec{q}}{2} \Big)\bra{\vec{p}-\vec{q}\,;1m_s',00}
t^{(1)}_{\gamma \eta}\ket{\vec{p}-
\vec{k}\,;1m_s,00}\psi_{m_s m_{d}}\Big(\vec{p}-\frac{\vec{k}}{2}  \Big),
\label{equ:t}
\end{eqnarray}
where $|1m_s,00\rangle$ denotes the two-nucleon spin and isospin wave 
function. 
Here, the intrinsic part of the deuteron wave function 
projected onto $|1m_s\rangle$ is denoted by 
\begin{equation}
\psi_{m_s m_{d}}(\vec p\,)=\sum_{l=0,2}\sum_{m_l}(l m_l 1 m_s|1 m_d) i^l u_l(p)
Y_{l,{m_l}}(\hat p)\,.
\end{equation}
For the radial wave functions $u_{0} $
and $u_{2} $, we have taken the ones of the Bonn 
r-space potential \cite{MaH87}. 

The operator $t_{\gamma \eta}^{(1)}$ is a
function of the photon, nucleon and eta momenta $\vec k$, $\vec p$, and 
$\vec q$, respectively, the photon polarization $\mu$, and the invariant 
mass $W_{\gamma N}$ of the photon-nucleon subsystem. As already mentioned 
in the introduction, implementing this operator into a bound system poses 
two problems. First of all, one has to know the invariant mass 
$W_{\gamma N}$ for the struck or active nucleon, and secondly, one needs 
the elementary $t$-matrix not in the $\gamma N$ c.m.\ system but in an 
arbitrary frame of reference. We will discuss 
these two points now in some detail. 

\subsection{ Choices for the invariant mass $W_{\gamma N}$ for the 
${\gamma N}$ subsystem}
For a bound system of two nucleons, the general expression for $W_{\gamma N}$ 
is, assuming the reaction to take place at nucleon ``1'', 
\begin{eqnarray}
W_{\gamma N} & = & \sqrt{(p_{10}+k_{0})^{2}-(\vec{p}_{1}+\vec k)^{2}} 
\nonumber \\
& = & \sqrt{(p_{10}^{\prime}+\omega_{q})^{2}-(\vec{p}_{1}\,^{\prime}+\vec 
q\,)^{2}},
\end{eqnarray}
where $p_{1\mu}^{(\prime)}=(p_{10}^{(\prime)},\vec p_1^{\,(\prime)})$ denotes 
its initial (final) four momentum. In general, one has $p_{10}^{(\prime)}
\not=\sqrt{M^{2}+\vec{p}\,^{\,(\prime)2}}$. In fact, the energy of an 
individual bound nucleon of a nonrelativistic many-particle system is not 
well defined. Only the total sum of the energies 
of all nucleons is a well defined quantity, e.g., for the deuteron 
$E^{d^{(\prime)}}=p_{10}^{(\prime)}+p_{20}^{(\prime)}$. 

One of many possible choices is to distribute the total energy of the deuteron 
equally on each of the two nucleons (Blankenbecler-Sugar choice) in the 
deuteron rest system, i.e., each nucleon has the energy $M_{d}/2$, 
independent of the momentum. Making such an assignment for the initial two 
nucleons, the boost to the $\gamma d$ c.m.\ system
gives then with $\vec p_1=\vec p - \vec k$ 
\begin{equation}\label{equ:w1}
W_{\gamma N}^{BS}=\frac{1}{E^{d}_k}\sqrt{\Big(\frac{1}{2}(E^{d}_k+k_{0})^2 
-\vec k \cdot \vec p\,\Big)^{2}-\vec{p}\,^{2}(E^{d}_k)^2} \,,
\end{equation}
while for the final two nucleons one finds 
\begin{equation}\label{equ:w1s}
W_{\gamma N}^{BS'}=\frac{1}{E^{d}_q}\sqrt{\Big(\frac{1}{2}(E^{d}_q+\omega_q)^2
-\vec q \cdot \vec p\,\Big)^{2}-\vec{p}\,^{2}(E^{d}_q)^2} \,.
\end{equation}

Another possibility is to take 
the active nucleon on-shell either before or after the interaction. In the 
first case, one finds with 
$p_{10}=E_{p-k}=\sqrt{M^{2}+(\vec{p}-\vec{k}\,)^{2}}$ 
\begin{equation}\label{equ:Ti}
W_{\gamma N}^{N}=\sqrt{(k_{0} +E_{p-k})^{2}-\vec{p}\,^{2}}\,,
\end{equation}
whereas in the second case with the final nucleon on-shell, i.e.\
$p_{10}^{\,\prime}=E_{p-q}=\sqrt{M^{2}+(\vec{p}-\vec{q}\,)^{2}}$, 
one has 
\begin{equation}\label{equ:finalN}
W_{\gamma N}^{N'}=\sqrt{(\omega_{q}+E_{p-q})^{2}-\vec{p}\,^{2}}\,.
\end{equation}
The former choice has been made in \cite{TiB94}. 
In all these three cases, the invariant mass depends on the 
kinematics, i.e., for fixed nucleon momentum on the angle between 
$\vec p$ and the outgoing $\eta$ momentum $\vec q$. 

The last choice, we want to discuss is to put the spectator nucleon on-shell, 
i.e., $p_{20}=E_{p}$. This choice has been taken in \cite{HoH79}. It 
is also used in coherent $\pi^0$ photoproduction on the deuteron \cite{WiA95}, 
and it may be justified by the fact that the deuteron is only loosely bound, 
and hence the spectator acts nearly like a free nucleon. We obtain in this case 
\begin{equation}\label{equ:eb}
 W_{\gamma N}^{S}=\sqrt{(W_{\gamma d}-E_{p})^{2}-\vec{p}\,^{2}}\,,
\end{equation} 
which is independent from the direction of $\vec p$.

Figure \ref{fig:1wsub} shows the invariant mass $W_{\gamma N}$ for these 
different choices as function of the spectator momentum at fixed 
photon lab energy $E_{\gamma}^{Lab}=800$ MeV. For the first four choices 
($W_{\gamma N}^{BS}$, $W_{\gamma N}^{BS'}$, $W_{\gamma N}^{N}$ 
and $W_{\gamma N}^{N'}$) we have represented the boundaries of the range 
spanned by the angle dependence by two curves corresponding to 
$\vec p$ and  $\vec k$ or $\vec q$ parallel (upper curve) and 
antiparallel (lower curve). 
One readily notes that $W_{\gamma N}^{N}$ spans the largest range, while 
the smaller ranges of $W_{\gamma N}^{BS}$, $W_{\gamma N}^{BS'}$ 
and $W_{\gamma N}^{N'}$ are compatible with each other. However, the average 
invariant masses nearly coincide 
for $W_{\gamma N}^{N}$ and $W_{\gamma N}^{N'}$, and they show a slight 
increase with increasing spectator momentum, whereas it decreases for both 
$W_{\gamma N}^{BS}$ and $W_{\gamma N}^{BS'}$ which, moreover, show a very 
similar behaviour. Finally, $W_{\gamma N}^{S}$ gives the lowest invariant 
mass with the strongest decrease with increasing $p$. 
One has to keep in mind that the main contribution to the total cross 
section originates from momenta 
$p$ below 400 MeV. But even in this region one notes a sizeable difference 
between the various choices for the invariant mass of the active $\gamma N$ 
subsystem. 

\subsection{Contribution of $\omega$ meson exchange in an arbitrary 
reference frame}
Now we will illustrate for the case of $\omega$ meson exchange the two 
methods for deriving the elementary production 
operator for a general frame of reference as mentioned above. To this end, 
we evaluate first the corresponding Feynman diagrams for an arbitrary 
frame (see the Appendix). The resulting $t$-matrix is expressed in terms 
of two amplitudes $\tilde M_v$ and $\tilde M_t$ which are also defined in 
the Appendix. These in turn can be represented as linear combinations of
the following operators
\begin{equation}
\omnull{a}{b}:=i \vec{\varepsilon}\cdot (\vec a \times \vec b)\,,
\hspace{1cm}
\Omega^1:= \vec{\sigma} \cdot \vec{\varepsilon}\,,\hspace{1cm}
\omzwei{a}{b}:= \vec{\sigma} \cdot \vec a\, \vec{\varepsilon} \cdot \vec b
\,.
\label{omoperator}
\end{equation} 
We then specialize to the $\gamma d$ c.m.\ system by the replacements 
$\vec p \rightarrow \vec p - \vec k$ and $\vec p^{\,\prime} 
\rightarrow \vec p - \vec q$, so that the $\gamma N$ subsystem moves with 
the total momentum $\vec p$. The corresponding coefficients 
for the amplitudes $\tilde M_v$ and $\tilde M_t$ are listed in 
Tab.\ \ref{tab:lt} where $s$, $t$ and $u$ stand for the Mandelstam 
variables of the $\gamma N$ subsystem. We further have introduced 
$p_0=\sqrt{\vec p^{\, 2}+s}$, the total energy of the subsystem, 
$E=\sqrt{(\vec p - \vec k\,)^2+M^2}$ and $E'=\sqrt{(\vec p - \vec q\,)^2+
M^2}$, the on-shell energies of the initial and final active nucleon, 
respectively, and $e_\pm= e'\pm e$ with $e^{(\prime)}= E^{(\prime)}+M$. 

Then, in order to proceed according to the second method, we specialize to the 
$\gamma N$ c.m.\ frame. Here one usually expresses the $t$-matrix in terms of 
the CGLN amplitudes. But we prefer to give them in terms of the operators 
defined in (\ref{omoperator}), 
i.e.\
\begin{eqnarray}
 \vec{\sigma} \cdot \vec{k}\,\vec{\varepsilon} \cdot \vec{q}
 &=& \omzwei{k}{q}\,,\\
 \vec{\sigma} \cdot \vec{q}\,\vec{\varepsilon} \cdot \vec{q}
 &=& \omzwei{q}{q}\,,\\
i\vec{\sigma} \cdot \vec{q}\,\vec{\sigma} \cdot (\vec{k}\times\vec{
\varepsilon}\,) &=& -\omnull{k}{q} -\vec k \cdot \vec q \,\omeins +
\omzwei{k}{q} \,.
\end{eqnarray}
We list the resulting coefficients in Tab.\ \ref{tab:cm}. 
These operators are 
then transformed to the $\gamma d$ c.m. frame by a proper Lorentz boost of 
all participating momenta, i.e., we boost $\vec k_{c.m.}$ and $\vec q_{c.m.}$ 
defined with respect to the $\gamma N$ c.m.\ frame to 
$\vec k$ and $\vec q$ with 
respect to the $\gamma d$ c.m. frame. In the latter frame the $\gamma N$ 
system moves with momentum $\vec p$ according to our choice of variables 
(see Fig.\ \ref{fig:kk}).
Then the Lorentz transformation reads
\begin{eqnarray}
\vec{k}_{c.m.} & = & \vec{k} +A_k\vec{p} 
\label{equ:d1}, \\
\vec{q}_{c.m.} & = & \vec{q} +A_q\vec{p} 
\label{equ:d2}\,,
\end{eqnarray}
with the boost parameter 
\begin{equation}\label{equ:boosta}
A_k =  \frac{1}{W_{\gamma N}}\Big( \frac{\vec{k}\cdot\vec{p}}
{E_{\gamma N}+W_{\gamma N}}-k_{0} \Big)\,, \\ 
\end{equation}
where $E_{\gamma N}=\sqrt{W_{\gamma N}^2+\vec p^{\,2}}$ denotes the energy 
of the $\gamma N$ subsystem, 
and $A_q$ is given by a corresponding expression 
replacing $k_\mu$ by $q_\mu$ in 
(\ref{equ:boosta}). Expressing $k_0$ and $\omega_q$ in terms of the 
invariant mass, the energy and the total momentum of the subsystem
\begin{eqnarray}
k_{0}&=& \frac{1}{2E_{\gamma N}}\Big(W_{\gamma N}^2 -M^2 
+2\vec{k}\cdot\vec{p}\Big)\,,\\
\omega_q&=& \frac{1}{2E_{\gamma N}}\Big(W_{\gamma N}^2 -M^2+m_\eta^2
+2\vec{q}\cdot\vec{p}\Big)\,,
\end{eqnarray}
we find for the boost parameters
\begin{eqnarray}
A_k & =&  -\frac{1}{2W_{\gamma N}E_{\gamma N}}\Big( W_{\gamma N}^2 
-M^2 +\frac{2W_{\gamma N}}
{E_{\gamma N}+W_{\gamma N}}\vec{k}\cdot\vec{p} \Big)\,, \\
A_q & =&  -\frac{1}{2W_{\gamma N}E_{\gamma N}}\Big( W_{\gamma N}^2+m_\eta^2 
-M^2 +\frac{2W_{\gamma N}}
{E_{\gamma N}+W_{\gamma N}}\vec{q}\cdot\vec{p} \Big)\,, 
\end{eqnarray}

For the resulting transformation of the operators we find 
\begin{eqnarray}
\omnull{k}{q} & \rightarrow & \omnull{k}{q} + A_k\,\omnull{p}{q} 
+ A_q\,\omnull{k}{p} \,,\\
\Omega^1 & \rightarrow & \Omega^1\,,
\\
\omzwei{k}{q} 
&\rightarrow &  \omzwei{k}{q} + A_k\,\omzwei{p}{q} + A_q\,\omzwei{k}{p} + 
A_k A_q\,\omzwei{p}{p} \,,\\
\omzwei{q}{q} 
&\rightarrow & \omzwei{q}{q} + A_q\,\omzwei{p}{q} + A_q\,\omzwei{q}{p} + 
 A_q^2\,\omzwei{p}{p} \,,
\end{eqnarray}
leading to the same
operator structures as for the general case. However, they differ in 
their corresponding coefficient functions as can already be seen by comparing 
the first four operators in Tab.\ \ref{tab:lt} with the corresponding ones 
in  Tab.\ \ref{tab:cm} which are not affected by the transformation. 
They differ just by terms which vanish in the 
$\gamma N$ c.m.\ frame. Thus it is not surprising that these terms cannot 
be generated by a boost. This means that information has been lost when going 
first to the $\gamma N$ c.m.\ frame with subsequent boost. Furthermore, for the 
remaining operators, which vanish in the $\gamma N$ c.m.\ frame, one finds 
little resemblance between the two methods. For example, the coefficient of 
$\omzwei{p}{p}$ in $\tilde M_v$ vanishes in Tab.\ \ref{tab:lt} 
whereas the one resulting from the 
transformation of $\omzwei{k}{q}$ and $\omzwei{q}{q}$ does not vanish. 
In Sect. \ref{kap:2} 
we will discuss the effect of these differences on the cross sections. 

\section{Results and Discussion}\label{kap:2}
Having fixed the parameters of the $\eta$ production model for the elementary 
process we can now proceed to study the coherent reaction on the deuteron. 
The $t$-matrix elements of (\ref{equ:t}) have been evaluated numerically 
using Gauss integration in momentum space. As deuteron wave functions, we 
have taken the ones of the Bonn r-space potential \cite{MaH87}. With 
respect to the elementary amplitudes, only the value for the helicity 
amplitude of the neutron $A^{n}_{1/2}$ remained undetermined. We have 
listed several values for $A^{n}_{1/2}$ in Table \ref{tab:game3} to be 
discussed in the following together with the resulting ratios of the 
neutron to proton amplitude $A^{n}_{1/2}/A^{p}_{1/2}$ and the correlated 
values for $e\kappa^{s}_{S_{11}}$ and the ratio of the isoscalar to the 
proton amplitude $A^{s}_{1/2}/A^{p}_{1/2}=1+\frac{A^{n}_{1/2}}{A^{p}_{1/2}}$. 
In addition, the last line of Table \ref{tab:game3} summarizes the range of 
values for these quantities which can be found in the literature 
(\cite{HaR89,KrA95b,PDG94} and references therein) and which are obtained 
either from photo reactions or from quark model predictions. 

\begin{figure}
\centerline{\psfig{figure=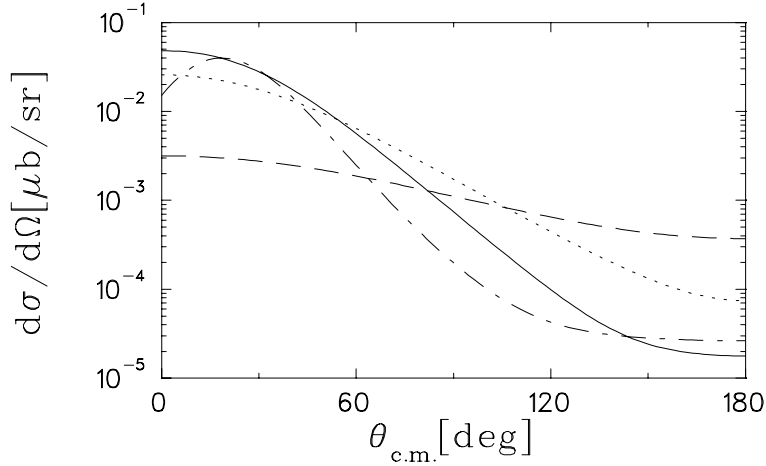,width=10cm,angle=0}}
\vspace*{1cm}
\caption{Differential cross sections for various photon energies 
using $A^{n}_{1/2}=-106\cdot10^{-3}$GeV$^{-\frac{1}{2}}$: 
dashed curve for $640$ MeV, dotted curve for $700$ MeV, full curve for $800$
MeV, and dashed-dotted curve for $E_{\gamma}^{Lab}=1000$ MeV.
\label{fig:4kurven}}
\end{figure}

First, we will consider the experimental results of Krusche et al.\ 
\cite{KrA95b} in order to discuss the limits on the neutron amplitude 
$A^{n}_{1/2}$ as imposed by present experimental data. 
Krusche et al.\ have measured the total and differential cross 
sections for quasifree $\eta$ photo production on the deuteron for energies 
from threshold up to 800 MeV, and they found for the ratio of neutron to
proton total cross sections a value $\frac{\sigma_{n}}{\sigma_{p}} 
\simeq \frac{2}{3}$. 
Furthermore, they have estimated from their data for the neutron resonance 
amplitude $A^{n}_{1/2}=-(100\pm30)\cdot 10^{-3}$GeV$^{-\frac{1}{2}}$ 
and for the cross section of the coherent process a value 
$(10^{-3}\pm 10^{-2})\,\mu$b/sr. This clearly 
indicates that the excitation of the $S_{11}$ resonance is dominated by the 
isovector amplitude. Consequently, the neutron and proton helicity 
amplitudes should have opposite sign, i.e., 
\begin{equation}\label{equ:an}
A^{n}_{1/2}  = \pm \sqrt{\frac{\sigma_{n}}{\sigma_{p}}}A^{p}_{1/2}\,.
\end{equation}
With our choice of $A^{p}_{1/2}=130\cdot 10^{-3}$GeV$^
{-\frac{1}{2}}$ (Table \ref{tab:a}) we get
$A^{n}_{1/2}=-106\cdot 10^{-3}$GeV$^{-\frac{1}{2}}$. It corresponds to the 
parameter set (A) in Table \ref{tab:game3}. 
The differential cross sections corresponding to this value of $A^{n}_{1/2}$ 
are shown in Fig.\ \ref{fig:4kurven} for four different photon lab energies 
$E_{\gamma}^{Lab}$. The results are well within 
the limit given by \cite{KrA95b} for the coherent production on the deuteron.

\begin{figure}
\centerline{\psfig{figure=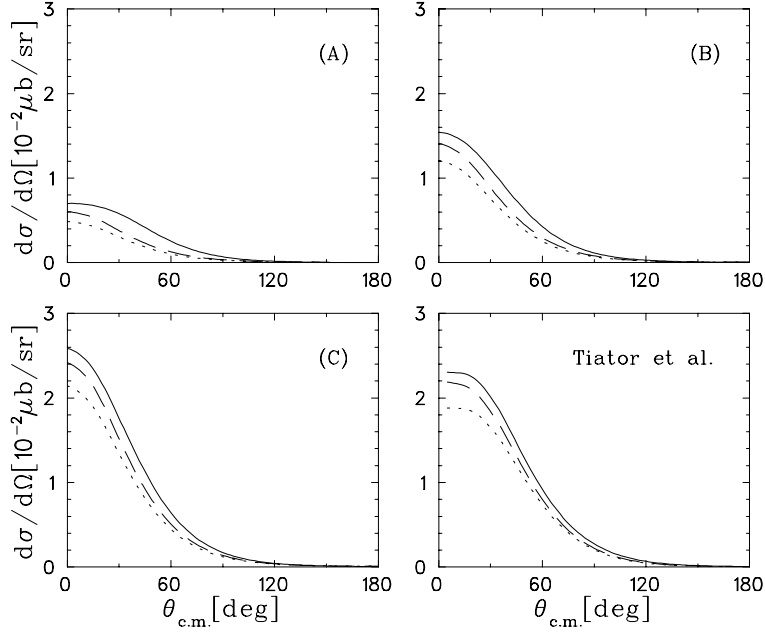,width=10cm,angle=0}}
\vspace*{1cm}
\caption{Differential cross sections at $E_{\gamma}^{Lab}=700$ MeV
for different choices of the neutron 
$S_{11}$ amplitude corresponding to the sets (A) to (C) of Table 
\protect{\ref{tab:game3}}.
The lower right panel shows the result of \protect{\cite{TiB94}}.
The the dashed curves include only the resonance, the
dotted ones the resonance and nucleon pole terms, and 
the full ones include all contributing graphs. 
\label{fig:kappa_s}}
\end{figure}

Fig.\ \ref{fig:kappa_s} shows the differential cross sections at a photon 
lab energy $E_{\gamma}^{Lab}=700$ MeV for the choices (A) through 
(C) for $A^{s}_{1/2}/A^{p}_{1/2}$ 
listed in Table \ref{tab:game3} and, for comparison, the result of Tiator et 
al.\ \cite{TiB94}. We have also indicated separately 
the contributions of the resonance and the nucleon Born terms. One 
readily sees the strong sensitivity on $\kappa^{s}_{S_{11}}$. An increase 
of $\kappa^{s}_{S_{11}}$ by a factor of two ((A) $\rightarrow$ (C)) leads 
to an increase of the cross 
section by almost a factor four. One also notes a destructive interference 
between resonance and nucleon pole term contributions whereas the vector 
meson exchange terms lead to a slight increase. Case (A) corresponds to the 
already discussed estimate of Krusche et al.\ while (B) is motivated by 
quark model predictions. Taking these as upper limit, then 
remains for the ratio 
$A^{s}_{1/2}/A^{p}_{1/2}$ only the value 
0.14 (set (B) in Table \ref{tab:game3}). 
The next case (C) serves for comparison with \cite{TiB94}.
If we take the same ratio for 
$A^{s}_{1/2}/A^{p}_{1/2}$ as in \cite{TiB94}, namely 0.15 which is very close 
to set (B) in Table \ref{tab:game3}, we get only 75 \% of their result. 
However, doubling the 
ratio of set (A), yielding set (C) of Table \ref{tab:game3}, we find 
a cross section that is qualitatively equal to the one of \cite{TiB94}.
However, our angular distribution drops 
somewhat faster with increasing angle than the one of \cite{TiB94}. 

The main differences between our calculation and the one of Tiator et al.\ 
\cite{TiB94} are 
that (i) they include in addition the resonances $P_{11}$ and $D_{13}$, but all 
resonances only in the $s$-channel, (ii) their elementary production model 
is a mixture of the isobar model and a coupled channel calculation, (iii) they 
start from the CGLN amplitudes in the $\gamma N$ c.m.\ frame and then boost 
all momenta as described above, (iv) they do not integrate over the 
spectator momentum but make use of the so-called factorization approximation 
which leads to uncertainties of about  5 to 10~$\%$ \cite{Kam95}, (v)  they 
have used an $\eta N$-coupling constant for the nucleon Born
terms of $g_{\eta}^{2}/4\pi=0.1$ in contrast to our value of 0.4, and (vi) 
they have chosen $W_{\gamma N}^{N'}$ as invariant mass of the $\gamma N$ 
subsystem.

\begin{figure}
\centerline{\psfig{figure=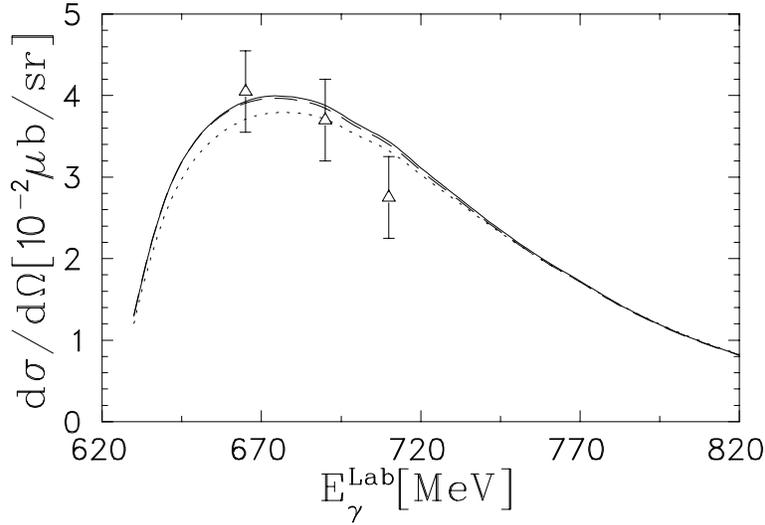,width=10cm,angle=0}}
\vspace*{1cm}
\caption{Differential cross section at $\theta_{c.m.} =90^{\circ}$ as 
function of the photon lab energy. The notation of the curves as in Fig.\
\protect{\ref{fig:kappa_s}}. Here we have used 
\protect{$e\kappa^{s}=340\cdot 10^{-3}$GeV$^{-\frac{1}{2}}$}. 
The data are taken from \protect{\cite{AnP69}}.
\label{fig:anp69}}
\end{figure}

The earlier experiment of Anderson and Prepost \cite{AnP69} seemed to 
indicate that the isoscalar
amplitude of the resonance is the dominant part instead of the isovector
one in contradiction to quark model predictions which give 
$A^{s}_{1/2}/A^{p}_{1/2} \simeq (-0.02 \pm 0.16)$
(see \cite{HaR89} and references therein). In fact, the calculations of
\cite{KrL76} in the impulse approximation require for this ratio a much 
larger value of 0.84 in order to reproduce the data of \cite{AnP69}. 
Even if one considers pion and eta rescattering terms 
one still needs a value of 0.6 \cite{HaR89}. 
In our approach, we could reproduce the data of \cite{AnP69} 
with a value of 0.88 corresponding to 
$A_{1/2}^{n}=114\cdot 10^{-3}$GeV$^{-\frac{1}{2}}$ and 
$e\kappa^{s}_{S_{11}}= 340\cdot 10^{-3}$GeV$^{-\frac{1}{2}}$ as is shown in 
Fig.\ \ref{fig:anp69}. On the other hand, the data of \cite{AnP69} are 
obviously at variance with the measurements of \cite{KrA95b} mentioned 
above which clearly prefer the isovector part to be the dominant one. 
This is also supported by comparing our results with the upper limit near
threshold obtained by Beulertz \cite{Beu94} which will be discussed in more
detail next. The cross section exceeds
this upper limit by more than a factor four. This fact and more recent 
experiments with a better background reduction imply
that very likely the experiment \cite{AnP69} has included other events
and therefore has resulted in an overestimation of the $\eta$ production cross 
section.

\begin{figure}
\centerline{\psfig{figure=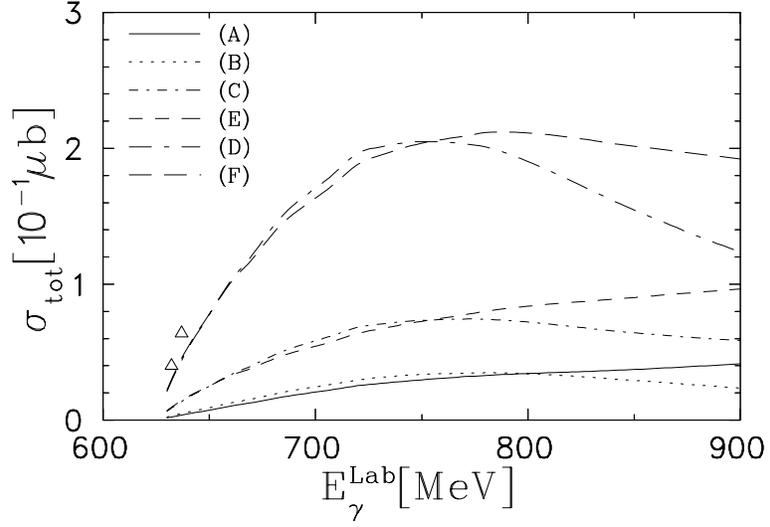,width=10cm,angle=0}}
\vspace*{1cm}
\caption{Total cross section for the six different values of the neutron 
amplitude (A) - (G) of Tab.\ \protect{\ref{tab:game3}}. 
The triangles represent the upper limits of \protect{\cite{Beu94}}.
\label{fig:game3}}
\end{figure}

\begin{figure}
\centerline{\psfig{figure=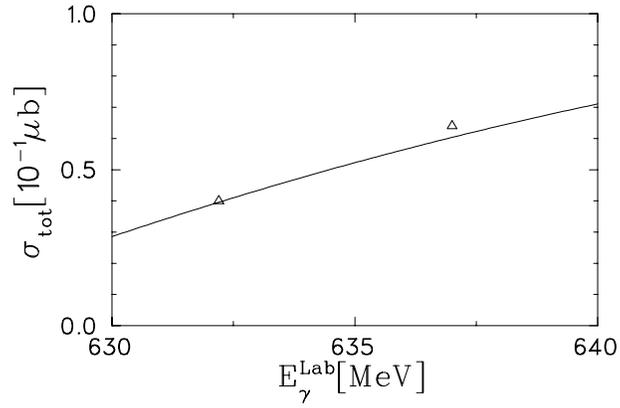,width=8cm,angle=0}}
\vspace*{1cm}
\caption{Total cross section for a coupling constant 
\protect{$e\kappa^{s}_{S_{11}}=140\cdot10^{-3}$GeV$^{-\frac{1}{2}}$}
(set (G) of Tab.\ \protect{\ref{tab:game3}}). 
The triangles represent the upper limits of \protect{\cite{Beu94}}.
\label{fig:schranke}}
\end{figure}

Now we will study the question for which  values of $A_{1/2}^{n}$ in Table 
\ref{tab:game3}
the resulting total  cross section is compatible with the upper limits 
obtained by Beulertz \cite{Beu94}. Instead of detecting the $\eta$ meson by 
its two-photon decay as, for example, is done by \cite{KrA95b}, the recoiling 
deuteron has been used as signal and 
the upper limits $\sigma_{tot}<0.040\, \mu$b for $E_{\gamma}^{Lab}=632.2$ 
MeV and $<0.064\,\mu$b for $E_{\gamma}^{Lab}=637$ MeV have been obtained. 
In Fig.\ \ref{fig:game3} we show theoretical total cross sections for the six 
different values of $A_{1/2}^{n}$ for the parameter sets (A) through (F) of 
Table \ref{tab:game3}. Although these resonance couplings vary over a wide 
range, the variation in the total cross sections is much less pronounced in 
the near threshold region than at higher energies. Even a sign change in 
$e\kappa^s_{S_{11}}$ does not show a big influence as one can see by 
comparing the results of 
set (C) with (E) or (D) with (F) except above the resonance region where 
the interference with the background becomes more important. The 
experimental upper limit is reached for the set (G) as is shown in Fig.\ 
\ref{fig:schranke}. 

All parameter sets of Table \ref{tab:game3} give total cross sections 
compatibel with the experimental estimates for the coherent deuteron cross 
section of \cite{Beu94} and \cite{KrA95b}. However, if one takes the 
quark model seriously, the parameter sets (D) through (G) can 
be excluded while (C) is on the borderline. 
Thus we conclude that the most probable parameter sets are (A) through (C) 
which give a neutron amplitude $A_{1/2}^{n}$ between $-82\cdot 10^{-3}$ and 
$-106\cdot 10^{-3}$GeV$^{-\frac{1}{2}}$.
The set (A) reproduces the experimental ratio for 
$\sigma_{n}/\sigma_{p}=2/3$ \cite{KrA95b}, whereas we find the best 
agreement with the theoretical results of \cite{TiB94} for the set (C). 
In this case, the ratio of neutron to proton cross section is only 2/5.

\begin{figure}
\centerline{\psfig{figure=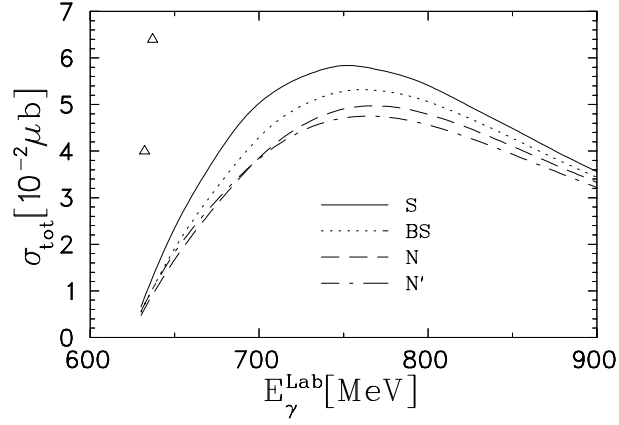,width=8cm,angle=0}}
\vspace*{1cm}
\caption{Total cross section for the coherent reaction on the deuteron 
including only the resonance $S_{11}$ for different choices of the invariant 
mass of the active subsystem using parameter set (C). 
The full curve with $W_{\gamma N}^{S}$, the dotted with $W_{\gamma N}^{BS}$,
the dashed with $W_{\gamma N}^{N}$, and the dash-dotted with 
$W_{\gamma N}^{N'}$. The triangles represent the upper limit of 
\protect{\cite{Beu94}}.
\label{fig:wsub0668}}
\end{figure}

Now we want to discuss the influence of different choices 
for the invariant mass of the $\gamma N$ subsystem. 
Figure \ref{fig:wsub0668} presents the total cross sections for the
$S_{11}$ resonance alone obtained with four of the different choices for the 
invariant mass $W_{\gamma N}$ as discussed in Sect.\ \ref{deuteron}. 
We did not consider $W_{\gamma N}^{BS'}$ because it is very similar to 
$W_{\gamma N}^{BS}$. 
Since for this question the exact value of $A^{n}_{1/2}$ is not relevant, 
we have used here $A^{n}_{1/2}=-82\cdot10^{-3}$GeV$^{-\frac{1}{2}}$. 
One readily notes considerable differences for the various prescriptions. 
The largest total cross section is obtained  with the spectator on shell 
$W_{\gamma N}^{S}$ having its maximum at 750 MeV. If one puts the active 
nucleon on shell, i.e., takes $W_{\gamma N}^{N}$ or $W_{\gamma N}^{N'}$ 
instead, the maximum 
is decreased by about 18\% and slightly shifted towards higher energies. 

\begin{figure}
\centerline{\psfig{figure=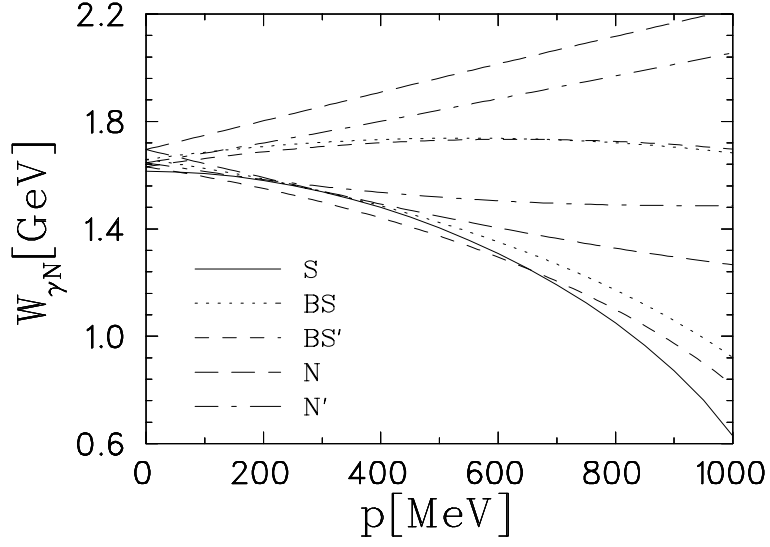,width=10cm,angle=0}}
\vspace*{1cm}
\caption{The invariant mass $W_{\gamma N}$ of the active photon-nucleon 
subsystem as a function of the spectator momentum $p$. The energy is 
$E_{\gamma}^{Lab}=800$ MeV, corresponding 
to $W_{\gamma d}=2553$ MeV. The full curve shows $W_{\gamma N}^{S}$. 
For each of the other invariant mass assignments $W_{\gamma N}^{BS}$, 
$W_{\gamma N}^{BS'}$, $W_{\gamma N}^{N}$, and $W_{\gamma N}^{N'}$, two 
curves show the borderlines of  the invariant mass region as explained in 
the inset. 
\label{fig:1wsub}}
\end{figure}

This decrease and shift can be understood as a result of the assignment 
of a higher invariant mass to the $\gamma N$ subsystem and the additional 
smearing due to the dependence on the angle between the spectator momentum 
and the photon, respective eta momentum (see Fig.\ \ref{fig:1wsub}) 
which leads to a larger effective width. The result is a slight upshift of 
the resonance position and a broadening, thus lowering of the maximum. 
One notes also little difference between putting the active nucleon before or 
after the interaction on shell. 
From the foregoing discussion it is apparent that the curve for the 
Blankenbecler-Sugar assignment $W_{\gamma N}^{BS}$ is about halfway 
between the spectator on shell and the active nucleon on shell because 
according to Fig.\ \ref{fig:1wsub}  $W_{\gamma N}^{BS}$ lies in between the 
spectator and active nucleon assignments. 

\begin{figure}
\centerline{\psfig{figure=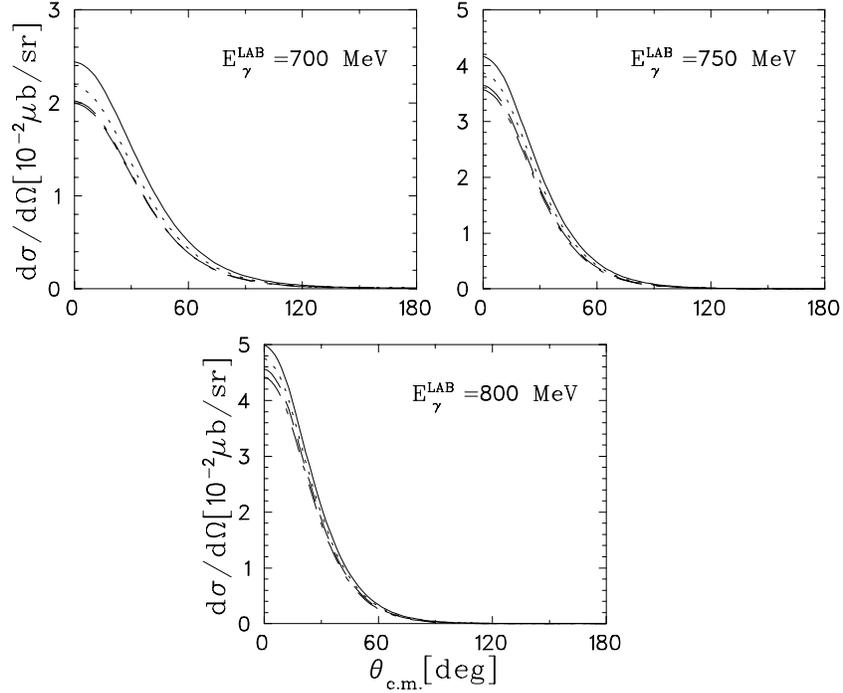,width=11cm,angle=0}}
\vspace*{1cm}
\caption{Differential cross sections at three lab photon energies 
including only the resonance $S_{11}$ for different choices of the invariant 
mass of the active subsystem using parameter set (C). 
The notation of the curves as in Fig.\ \protect{\ref{fig:wsub0668}}. 
\label{fig:difsub}}
\end{figure}

The differential cross sections for three photon energies in 
Fig.\ \ref{fig:difsub} show a similar behaviour with respect to the 
invariant mass assignment. In particular at lower energies (see as example 
the cross section at $E_\gamma=700$ MeV in Fig.\ \ref{fig:difsub}) the 
differences are quite sizeable in forward direction. 

In view of these results, we have to conclude that the choice for the invariant 
mass of the $\gamma N$ subsystem has a significant influence 
on the cross section. In order to obtain a cross section of the same magnitude 
for two different choices of $W_{\gamma N}$, one has to
change the elementary helicity amplitude, too. Consequently, the other
correlated parameters (see Table \ref{tab:game3}) will change together with 
the assignment for the invariant mass. This has also some bearing on the 
question of compatibility with quark model predictions.

As last point we want to discuss the different methods of deriving 
the elementary 
production operator in an arbitrary frame which is necessary for implementing 
it into the bound system, i.e., we will compare the 
more general case using an arbitrary frame (GC), the Lorentz boost of the 
momenta from the $\gamma N$ c.m.\ frame (LB), and the simplest approach of 
taking the elementary operator without any transformation of the variables 
(CM). The last method, for 
example, is considered in \cite{Wil92} for the coherent $\pi^0$ 
photoproduction on the deuteron where
only the momentum of the $\Delta$ resonance is transformed while for 
the background terms this has not been done, assuming the effect to 
be negligible. While for the pion this might be justified because of its 
low mass, one certainly would not expect it to be a valid approximation for 
the eta meson. 
  
\begin{figure}
\centerline{\psfig{figure=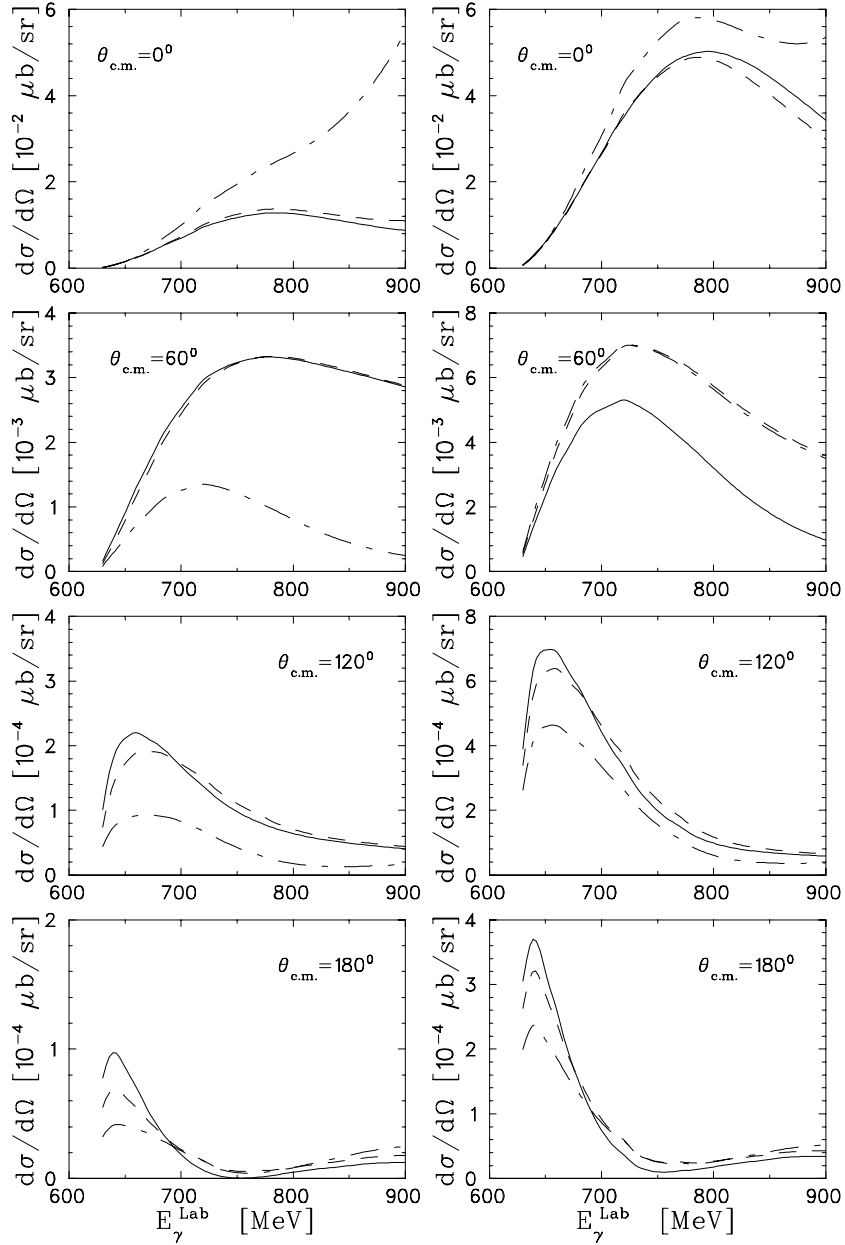,width=11cm,angle=0}}
\vspace*{1cm}
\caption{Differential cross section as function of the photon lab energy 
for various constant angles, 
on the left side for parameter set (A) and on the right side for parameter 
set (C). The full curve shows the calculation using the general expression 
for $\omega$ exchange (GC), the dashed one is obtained with the 
Lorentz boosted operator for the $\omega$ meson contribution (LB) and for the 
dashed-dotted curves the c.m.\ elementary operator has been used without any 
transformation (CM).
\label{fig:3maldeut22}}
\end{figure}

We present results for two choices of the neutron helicity amplitude, namely 
$A^{n}_{1/2}=-106\cdot10^{-3}$GeV$^{-\frac{1}{2}}$ (A) and
$A^{n}_{1/2}=-82\cdot10^{-3}$GeV$^{-\frac{1}{2}}$ (C) (Table
\ref{tab:game3}). The differential cross sections as a
function of the photon laboratory energy and for constant c.m.\ angles are 
shown for both amplitudes in Figs.\ \ref{fig:3maldeut22}.
It can be seen that the differences between (GC), (LB), and (CM) 
depend not only on the angles and energies but also 
on the strength of the helicity amplitude chosen. 

Comparing first (GC) and (LB), we see that at forward angles the 
deviations are small. However, for backward angles 
they become more significant reaching a maximum at 180$^{\circ}$. 
The maximum is considerably lower for (LB) while at higher energies the 
differential cross setion for (LB) lies above the one for (GC). 
On the other hand, the main contributions to the cross sections stem from
small angles where also the differences are smaller so that the total cross 
sections differ much less. Furthermore, using a larger helicity
amplitude $A^{n}_{1/2}$ decreases the differences between the calculations 
(GC) and (LB) even more. That is the case for an $A^{n}_{1/2}$ with which we get a cross
section comparable to the one of \cite{TiB94}.

Finally, comparing the calculation (CM), to which only the first four operator
structures of Table \ref{tab:lt} contribute, with (GC) one clearly sees very 
large deviations from (GC) which obviously are 
not tolerable.

\section{Summary and Conclusions}
Coherent eta photoproduction on the deuteron has been studied in 
the impulse approximation neglecting rescattering and two-body effects. 
For the elementary reaction, we have assumed that the process is dominated by 
the excitation of the $S_{11}$(1535) resonance 
for photon energies not too far above 
threshold. Thus we have included the $S_{11}$ in the $s$- and $u$-channels. 
In addition, we have considered as background the $t$-channel 
vector meson exchange with $\rho$ and $\omega$ as well as the nucleon 
pole terms in $s$- and $u$-channel for which pseudoscalar coupling has been 
assumed. The vertices are taken from an effective Lagrangian theory. 
The elementary process is treated nonrelativistically. All parameters are 
fixed by fitting the experimental data for the reaction on the proton. 

The electromagnetic coupling constant $A_{1/2}^{n}$ for the $S_{11}$ 
excitation on the neutron cannnot be determined this way, and one needs 
data for the reaction on the deuteron. In view of the fact that up to now 
no precise data on the deuteron are available, we could only use the 
existing estimates to find limits for this amplitude. The experiment for 
quasifree production 
\cite{KrA95b} gave a value for the ratio of neutron to proton cross section 
with which we have obtained a deuteron cross section that is within the 
experimentally obtained limits of \cite{Beu94} and \cite{KrA95b}.
Correlated with this is a ratio of the isoscalar to isovector amplitude  
that is consistent with quark model predictions. However, the upper limits 
of \cite{Beu94} and \cite{KrA95b} for the coherent deuteron 
cross section are not sufficient for a precise determination of $A_{1/2
}^{n}$. An uncertainty of the factor ten is to be considered which could 
result in deviations from quark model predictions up to 50\%. Furthermore, 
we have studied the theoretical problems of implementing the elementary 
reaction amplitude into a bound system connected to the choice of the 
invariant mass for the resonance excitation and to the transformation of 
the reaction amplitude from the $\gamma N$ c.m.\ frame to the c.m.\ frame 
of the $\gamma$-bound system, and we have found that the theoretical results 
show significant differences when using different prescriptions. 

For the future we need on the experimental side precise data for both the 
coherent and the incoherent production on the deuteron in order to be able 
to fix in a reliable manner the $S_{11}$ excitation on the neutron. On the 
theoretical side, we have to improve the treatment by including 
rescattering and two-body effects in order to obtain a more realistic 
description of this important process. 
 
\renewcommand{\theequation}{A.\arabic{equation}}
\setcounter{equation}{0}
\section*{Appendix: $t$-matrix contribution from $\omega$ meson exchange}
\label{kap:A2}

With the aid of the Lagrangians (\ref{lvnn})-(\ref{equ:he}) 
we get for the $t$-matrix contribution from vector meson exchange 
\begin{equation}\label{equ:a41}
i \tau_{fi}^{V}= \frac{e \lambda}{m_{\eta}} \frac{\varepsilon_{\mu\nu
\lambda\sigma}k^{\nu}\varepsilon^{\mu}q^{\lambda}}{t-m_{V}^{2}}
\bar{u}_{f}(p^{\prime}) \left[ g_{v} \gamma^{\sigma}- \frac{g_{t}}{2M }i 
\sigma^{\sigma\alpha}(q-k)_{\alpha} \right] u_{i}(p)\,.
\end{equation}
For the $\rho$ meson one has to multiply with an additional factor $\tau_{0}$.
In the CGLN-representation \cite{ChG57}, on has for photo production 
only four independent amplitudes 
\begin{equation}\label{equ:a1}
i\tau_{fi}=\bar{u}_{f}(p^{\prime}) \sum_{j=1}^{4}A_{j}(s,t,u)M_{j}u_{i}
(p)\,.
\end{equation}
containing the Dirac operators 
\begin{eqnarray}
M_{1}& = & -\frac{1}{2} \gamma_{5} \gamma_{\mu} \gamma_{\nu} 
(\epsilon^{\mu}k^{\nu}-\epsilon^{\nu}k^{\mu})
\,, \\
M_{2} & = &  \gamma_{5} (p+p^{\prime})_{\mu} \left (q_{\nu}-
\frac{1}{2}k_{\nu} \right )(\epsilon^{\mu}k^{\nu}-\epsilon^{\nu}k^{\mu}) \,, \\
M_{3} & = & - \gamma_{5} \gamma_{\mu} q_{\nu} 
(\epsilon^{\mu}k^{\nu}-\epsilon^{\nu}k^{\mu})\,, \\
M_{4} & = & - \gamma_{5} \gamma_{\mu} (p+p^{\prime})_{\nu}
(\epsilon^{\mu}k^{\nu}-\epsilon^{\nu}k^{\mu})-2M M_{1}\,, 
\end{eqnarray}
and the invariant amplitudes $A_j$ for which one finds from (\ref{equ:a41})
\begin{eqnarray} \label{equ:a111}
A_{1} & = & \frac{e \lambda}{m_{\eta}} \frac{g_{t}}{2M } \frac{t}{t-m_{V}
^{2}}\,, \\
A_{2} & = & -\frac{e \lambda}{m_{\eta}} \frac{g_{t}}{2M }\frac{1}{t-m_{V}
^{2}}\,,\\ \label{equ:a2}
A_{3} & = & 0\,, \\ \label{equ:a3}
A_{4} & = & -\frac{e \lambda}{m_{\eta}} g_{v}\frac{1}{t-m_{V}^{2}}\,. 
\label{equ:a4}
\end{eqnarray}

Introducing the operators $\tilde M_j$ in Pauli spinor space by 
$\chi^{\dag}_{f}\tilde{M_{j}}\chi_{i}:=\bar{u}_{f} M_{j} u_{i}$ and noting 
that $A_{3}=0$ for vector mesons, one finds for the remaining operators 
\begin{eqnarray}
\tilde{M_{1}} = N^{\prime}N \Big[ &&- k_{0} \vec{\sigma} \cdot
\vec{\varepsilon}- \frac{i \vec{\sigma} \cdot (\vec{k} \times \vec{
\varepsilon})\, \vec{\sigma} \cdot \vec{p}}{e_{p}}+\frac{i \vec{
\sigma}
\cdot \vec{p}\,^{\prime}\, \vec{\sigma} \cdot (\vec{k} \times \vec
{\varepsilon})}{e_{p^{\prime}}} \nonumber \\
 & &  +\frac{k_{0}}{e_{p}e_{p^{\prime}}} \vec{\sigma} 
\cdot \vec{p}\,^{\prime}
(\vec{p} \cdot \vec{\varepsilon}-i \vec{\sigma} \cdot (\vec{p} 
\times\vec{\varepsilon})) \Big] \,,\\
\tilde{M_{2}}   =   N^{\prime}N \Big[&& 
\frac{\vec{\sigma} \cdot \vec{p}\,^{\prime}}{e_{p^{\prime}}}
-\frac{\vec{\sigma} \cdot \vec{p}}{e_{p}}\Big]
\Big[ J\vec{\varepsilon}\cdot (\vec{p}\,^{\prime} +\vec{p}\,)
-K \vec{\varepsilon}\cdot \vec q\,\Big]\,,\\
\tilde{M_{4}}  =  N^{\prime}N \Big[ &&\left (k_{0} \frac{ \vec{
\sigma} \cdot \vec{p}\,^{\prime}}{e_{p^{\prime}}}+k_{0} \frac{ \vec{
\sigma} \cdot \vec{p}}
{e_{p}}- \vec{\sigma} \cdot \vec{k}- \vec{\sigma} \cdot \vec{p}\,^{\prime}
\, \vec{\sigma} \cdot \vec{k} \,\vec{\sigma}
\cdot \vec{p} \frac{1}{e_{p}e_{p^{\prime}}}\right)
 \nonumber  \\
& &  \times (\vec{p} \cdot
\vec{\varepsilon}+ \vec{p}\,^{\prime} \cdot \vec{\varepsilon})
 - \frac{K}{e_{p^{\prime}}e_{p}} \vec{\sigma} \cdot \vec{p}\,
^{\prime}
\, \vec{\sigma}  \cdot \vec{\varepsilon} \,\vec{\sigma} \cdot \vec{p}
- \vec{\sigma} \cdot \vec{\varepsilon} K\Big] 
-2M  \tilde{M_{1}}\,,
\end{eqnarray}
with $J = q\cdot k$, $K = (p+p^{\prime})\cdot k$, 
$N^{(\prime)}=\sqrt{ \frac{e_{p^{(\prime)}} }{2M }}$ and 
$e_{p^{(\prime)}} =E_{p^{(\prime)}}+M $. 

We will now bring (\ref{equ:a1}) into the form
\begin{equation}
i\tau_{fi}= \frac{e \lambda}{m_{\eta}} \frac{1}{t-m_{V}^{2}}\,
\frac{N'N}{e_{p'}e_{p}}\,
\chi^{\dag}_{f}\Big(-g_{v} \tilde{M}_{v} + \frac{g_{t}}{2M } 
\tilde{M}_{t}\Big)\chi_{i}\,,
\end{equation}
where we have introduced
\begin{eqnarray}
\tilde{M}_{v}&=&\frac{e_{p'}e_{p}}{N'N}\,\tilde{M}_{4}\,,\\
\tilde{M}_{t}&=&\frac{e_{p'}e_{p}}{N'N}\,(t\tilde{M}_{1}-\tilde{M}_{2})\,.
\end{eqnarray}
In terms of the operators introduced in (\ref{omoperator}), we find 
explicitly
\begin{eqnarray}
\tilde{M}_{v}&=&
\frac{1}{4}\Big( (s-u + 4Mk_0)t +2Me_-(m_\eta^2-t)\Big)
\Omega^1 +(\frac{t}{2} + E_pe_+)\Omega^0_{\vec k,\,\vec p^{\,\prime}}
\nonumber\\
&& -(\frac{t}{2} + E_{p'}e_+)\Omega^0_{\vec k,\,\vec p}
+k_0e_+\Omega^0_{\vec p^{\,\prime},\,\vec p}
+(\frac{t}{2}+Me_-)\Omega^2_{\vec k,\,\vec p}
+(\frac{t}{2}-Me_-)\Omega^2_{\vec k,\,\vec p^{\,\prime}}
\nonumber\\
&&+(k_0M + \frac{1}{2}(s-M^2) )(\Omega^2_{\vec p^{\,\prime},\,\vec p^{\,\prime}}-
\Omega^2_{\vec p,\,\vec p^{\,\prime}})
+(k_0M - \frac{1}{2}(u-M^2))(\Omega^2_{\vec p,\,\vec p}-
\Omega^2_{\vec p^{\,\prime},\,\vec p})\,,
\\
\tilde{M}_{t}&=&
-\frac{t}{2}\Big( k_0t+e_{p'}(s-M^2)-e_{p}(u-M^2)\Big)\Omega^1
+e_{p'}t\Big( \Omega^0_{\vec k,\,\vec p}-\Omega^2_{\vec k,\,\vec p}\Big)
\nonumber\\
&&-e_{p}t\Big( \Omega^0_{\vec k,\,\vec p^{\,\prime}}
  +\Omega^2_{\vec k,\,\vec p^{\,\prime}}\Big)
  -k_0t\Omega^0_{\vec p^{\,\prime},\,\vec p} 
  -e_{p}(s-M^2)\Omega^2_{\vec p^{\,\prime},\,\vec p^{\,\prime}}
  +e_{p'}(u-M^2)\Omega^2_{\vec p,\,\vec p}
\nonumber\\
&&+(k_0t + e'\,(s-M^2) )\Omega^2_{\vec p,\,\vec p^{\,\prime}}
+(k_0t - e\,(u-M^2))\Omega^2_{\vec p^{\,\prime},\,\vec p}\,.
\end{eqnarray}
Here, $s$, $t$ and $u$ denote the usual Mandelstam variables and 
$e_{\pm}=e_{p'}\pm e_{p}$. 

For the specialization to the $\gamma N$ c.m.\ sytem, we note the following 
relations using $\vec p = -\vec k$ and $\vec p^{\,\prime} = -\vec q$
\begin{eqnarray}
k_0&=&\frac{1}{2W}(s-M^2)\,,\\
e_p&=&\frac{1}{2W}(W+M)^2\,,\\
e_{p'}&=&\frac{1}{2W}((W+M)^2-m_\eta^2)\,,
\end{eqnarray}
with $W^2=s$.

\begin{table}
\caption{Parameters for the vector mesons.}
\begin{tabular}{rccccl}
 & $m_{V}$ (MeV) & $\lambda$ & $\lambda g_{t}$ & $\lambda g_{v}$ & $\Lambda_{V}
$ (MeV) \\
\hline
$\rho$ & 770 & 0.890 & 13.6085 & 2.2309 & 1800 \\
$\omega$ & 782 & 0.192 & 0 & 3.2640 & 1400 \\
\end{tabular}
\label{tab:d}
\end{table} 

\begin{table}
\caption{Resonance parameters giving the best agreement with the data.}
\begin{tabular}{ccccc}
$\Gamma^{0}_{S_{11}}$ (MeV) & $M_{S_{11}}$ (MeV) & 
$A_{1/2}^{p}$ (GeV$^{-\frac{1}{2}}$) & $g_{\eta NS_{11}}$ & $g_{\pi NS_{11}}$ \\
\hline
150 & 1535 & $130 \cdot 10^{-3}$ & 2.033 & 1.148 \\
\end{tabular}
\label{tab:a}
\end{table}

\begin{table}
\caption{
Contribution of $\omega$-meson  to the $t$-matrix in the 
the $\gamma d$ c.m.\ system. 
}
\begin{flushleft}
\begin{tabular}{c||c|c}
  operator 
& $\tilde M_{v} $ 
& $\tilde M_{t}$ 
\\
\hline
  $\Omega^1$
& $ \frac{1}{4}\Big( (s-u + 4Mk_0)t +2Me_-(m_\eta^2-t)\Big)$
& $ -\frac{t}{4}\Big( 2k_0t+e_+(s-u)+e_-(m_\eta^2-t)\Big)$
\\
\hline
  $\Omega^0_{\vec k,\,\vec q}$
& $ -(\frac{t}{2}+p_0e_+)$
& $ (p_0 + M)t$
\\
\hline
  $\Omega^2_{\vec k,\,\vec q}$
& $\frac{1}{2}(u-M^2-m_\eta^2)+M(e_--k_0)$
& $(p_0 + M)t+e'(s-M^2)$
\\
\hline
  $\Omega^2_{\vec q,\,\vec q}$
& $\frac{1}{2}(s-M^2)+Mk_0$
& $-e(s-M^2)$
\\
\hline
\hline
  $\Omega^0_{\vec q,\,\vec p}$
& $-k_0e_+$
& $k_0t$
\\
\hline
  $\Omega^2_{\vec q,\,\vec p}$
& $ \frac{1}{2}(t-m_\eta^2)$
& $ e\,m_\eta^2-(k_0+e)t$
\\
\hline
  $\Omega^0_{\vec k,\,\vec p}$
& $e_+(k_0-e_-)$
& $(e_--k_0)t$
\\
\hline
  $\Omega^2_{\vec k,\,\vec p}$
& $\frac{1}{2}(t+m_\eta^2)$
& $-\Big( (p_0+M)t+e'm_\eta^2\Big)$
\\
\hline
  $\Omega^2_{\vec p,\,\vec q}$
& $$
& $-\Big(k_0t+e_-(s-M^2)\Big)$
\\
\hline
  $\Omega^2_{\vec p,\,\vec p}$
& $$
& $(2k_0-e_-)t +e_-m_\eta^2$
\\
\end{tabular}
\end{flushleft}
\label{tab:lt}
\end{table}

\begin{table}
\caption{Contribution of \protect{$\omega$}-meson  to the $t$-matrix in the 
$\gamma N$ c.m.\ system. The invariant mass is denoted by 
$W=W_{\gamma N}$ ($W^2=s$). 
}
\begin{flushleft}
\begin{tabular}{c||c|c}
  operator 
& $\tilde M_{v} $ 
& $\tilde M_{t}$ 
\\
\hline
  $\Omega^1$
& $ \frac{1}{4}\Big( (s-u + \frac{2M}{W}(s-M^2))t -\frac{m_\eta^2 M}{W}
(m_\eta^2-t)\Big)$
& $ -\frac{t}{4W}\Big( t(s-M^2)-\frac{1}{2}m_\eta^2(m_\eta^2-t)$
\\
& 
& $+((W+M)^2-\frac{1}{2}m_\eta^2)(s-u)\Big)$
\\
\hline
  $\Omega^0_{\vec k,\,\vec q}$
& $ -\Big(\frac{1}{2}(t-m_\eta^2) +(W+M)^2\Big)$
& $ (W + M)t$
\\
\hline
  $\Omega^2_{\vec k,\,\vec q}$
& $\frac{1}{2}\Big(u-M^2-m_\eta^2-\frac{M}{W}(s-M^2-m_\eta^2)\Big)$
& $(W + M)t+\frac{1}{2W}((W+M)^2-m_\eta^2)(s-M^2)$
\\
\hline
  $\Omega^2_{\vec q,\,\vec q}$
& $\frac{1}{2}(s-M^2)(1+\frac{M}{W})$
& $-\frac{(W+M)^2}{2W}(s-M^2)$
\\
\end{tabular}
\end{flushleft}
\label{tab:cm}
\end{table}

\begin{table}
\caption{ Various values for $A^{n}_{1/2}$, the ratio $A^{n}_{1/2}/A^{p}_{1/2}$
and the correlated values for $e\kappa^{s}_{S_{11}}$  and 
$A^{s}_{1/2}/A^{p}_{1/2}$ as used in 
this work. In addition, the last line shows the range of values that is 
found in the literature (see text).}
\begin{tabular}{|c|cccc|}
 & $A^{n}_{1/2}\,[10^{-3}$GeV$^{-\frac{1}{2}}]$ & $A^{n}_{1/2}/A^{p}_{1/2}$ 
& $e\kappa^{s}_{S_{11}}\, [10^{-3}$GeV$^{-\frac{1}{2}}]$ & 
$A^{s}_{1/2}/A^{p}_{1/2}$ \\
\hline
(A) & -106 & -0.82 &  33.4   &  0.09 \\
(B) & - 93 & -0.72 &  51.0   &  0.14 \\
(C) & - 82 & -0.63 &  66.8   &  0.18 \\
(D) & - 43 & -0.33 & 120.6   &  0.33 \\
(E) & -178 & -1.37 & -\,66.8 & -0.18 \\
(F) & -217 & -1.67 & -120.6  & -0.33 \\
(G) & -29  & -0.23 &  140.0  & 0.39 \\ 
\hline
(Lit.) & -11 to -119 & -0.65 to -1.5 &  & -0.18 to +0.14 \\ 
\end{tabular}
\label{tab:game3}
\end{table}

\end{document}